\input epsf
\documentstyle[12pt,fleqn]{article}
\def\be{\begin{equation}}
\def\ee{\end{equation}}
\def\ba{\begin{eqnarray}}
\def\ea{\end{eqnarray}}

\def\fun#1#2{\lower3.6pt\vbox{\baselineskip0pt\lineskip.9pt
        \ialign{$\mathsurround=0pt#1\hfill##\hfil$\crcr#2\crcr\sim\crcr}}}

\begin{document}
\begin{titlepage}
\null\vspace{-62pt}

\begin{flushright}
UCLA/97/TEP/20\\
IEM-FT-163/97\\
hep-ph/9709289\\
Revised Version April 98\\
\end{flushright}
\vspace{0.3in}
\begin{center}{\Large\bf Generation of Large Lepton 
Asymmetries }
\footnote{Research supported in part by: the CICYT
(contract AEN95-0195) and the European Union
(contract CHRX-CT92-0004) (JAC); 
and the US Department of Energy
under grant DE-FG03-91ER40662 Task C (GG).}
\\[2cm]
\end{center}
\begin{center}
  {\large\bf Alberto Casas}$^{\mbox{\small a}}$,
  {\large\bf Wai Yan Cheng}$^{\mbox{\small b}}$, 
  {\large\bf  Graciela Gelmini}$^{\mbox{\small c}}$\footnote{On leave at CERN (Theory Division) until June 15, 1998.}\\[2ex]
 \end{center}

\begin{center}
$^{\mbox{\small a}}${Instituto de Estructura de la Materia, CSIC\\
 Serrano 123, 28006 Madrid, Spain\\
 casas@cc.csic.es}

\vspace{0.5cm} $^{\mbox{\small b}}$$^{\mbox{\small c}}${Department 
of Physics and Astronomy,
  University of California, UCLA\\
  405 Hilgard Ave., Los Angeles, CA 90095\\
 $^{\mbox{\small b}}$wcheng@physics.ucla.edu,\\ 
 $^{\mbox{\small c}}$gelmini@physics.ucla.edu}
 \end{center}
\vspace{2cm}

\begin{abstract}
We present here a realistic model to produce a very large lepton
asymmetry $L\simeq 10^{-2} - 1$, without producing a large baryon asymmetry.
The model is based on the Affleck and Dyne scenario, in which the field
acquiring a large vacuum expectation value during inflation is a s-neutrino
right. We require $L$ to be large enough for the
 electroweak symmetry  to be spontaneously broken at all temperatures after 
 inflation, with the consequent suppression of sphaleron transitions.
\end{abstract}

\end{titlepage}



\baselineskip=24pt

\vspace{24pt}

\newcommand{\gsim}
{\mathrel{\raise.3ex\hbox{$>$\kern-.75em\lower1ex\hbox{$\sim$}}}}
\newcommand{\lesssim}
{\mathrel{\raise.3ex\hbox{$<$\kern-.75em\lower1ex\hbox{$\sim$}}}}

\newpage

The possible role of neutrino degeneracy in nucleosynthesis has been studied
several times from 1967 onwards \cite{nucleo} \cite{ks}. 
 A large number of electron neutrinos, $\nu_e$, present
during nucleosynthesis yields a reduction of the neutron to proton ratio 
$n/p$, through the reaction $n ~\nu_e \to p~e$. This in turn lowers
the ${}^4He$ abundance, since when nucleosynthesis
takes place essentially all neutrons end up in  ${}^4He$ nuclei.
 Extra neutrinos of any flavor would
increase the energy density of the Universe, leading to
 an earlier decoupling of weak interactions and consequent increase of $n/p$
(and  ${}^4He$). This last effect is less important than the former one
 in the case of electron neutrinos, but it is the only 
effect of an excess of muon and/or tau neutrinos,
$\nu_\mu$ and $\nu_\tau$.  Thus when both the chemical potentials of
${\nu_e}$  and of $\nu_{\mu}$, $\nu_{\tau}$
are large, their effect largely compensate each other.  This allows to obtain
light element abundances in agreement with observations even for values of
the baryon to photon ratio
$\eta\equiv n_B/n_\gamma$ much larger than in standard nucleosynthesis, which
is $\eta \simeq O(10^{-10})$.
Therefore, the usually quoted nucleosynthesis upper bound on the cosmological
density of baryons $\Omega_B$ may be falsified and much larger values of
$\Omega_B$ become allowed by nucleosynthesis, even $\Omega_B=1$
 (which corresponds to $\eta \simeq O(10^{-8})$).  
 
 More restrictive upper bounds on $\Omega_B$ can be obtained,
however, by examining other consequences of a large lepton number 
\cite{fkt}.  Due to
the electric charge neutrality of the Universe, the excess of protons with
respect to antiprotons must be accompanied by the same excess of electrons over
positrons, so
$L_{e\pm} \equiv (n_e - n_{\bar e})/s$ $\simeq B \equiv (n_B -n_{\bar B})/s 
\simeq O(10^{-10}-10^{-8})$.  A larger lepton
number can reside only in neutrinos. 
 We will be interested in values of $L =$
$L_e + L_\mu + L_\tau$ of order one.  Thus to a very high accuracy
$L_\alpha = ({n_{\nu_\alpha} - n_{\bar\nu_\alpha}) / s}$, where
$\alpha = e, \mu, \tau$.  Here $n$ stands
 for the number density of the particle indicated in the suffix
and $s$ is the entropy density of the Universe. The asymmetries
$L_\alpha$ are related to the chemical potentials, $\mu_\alpha$,
\begin{equation}
\label{asymm}
L_\alpha = {45 \over 12 \pi^4 
g_{s*}(T_\gamma)} \left ({T_\nu \over T_\gamma}\right )^3
\left (\pi^2 \xi_\alpha + \xi^3_\alpha\right ) \simeq 3.6
 \left (10^{-2} \xi_\alpha + 10^{-3} \xi_\alpha^3\right )~,
\end{equation}
where $\xi_\alpha \equiv \mu_{\nu _\alpha}/T_\nu$ 
 are dimensionless chemical potentials, $g_{s*}$ is
the $T_\gamma$ dependent entropy number of degrees of freedom, 
i.e. $s =(2 \pi^2 / 45) g_{s*} T_\gamma^3=$
$\pi^4 g_{s*} n_{\gamma} / 45 \zeta(3) \simeq$ $1.8008 g_{s*} n_{\gamma}$, 
and the numerical relation holds during nucleosynthesis. Notice that, 
with only the relativistic particles in the standard model 
present during nucleosynthesis, the relation between the temperature of
neutrinos, $T_\nu$, and the temperature of photons, $T_\gamma$, is given
by $g_{s*} \left ( {T_\gamma / T_\nu}\right )^3 = 10.75$, 
both before and after $e^+ e^-$ annihilation, what means 
$T_\nu = T_\gamma$ before $e^+ e^-$ annihilation. In 
the presence of large neutrino asymmetries this relation holds only
if $\xi_\alpha < 12 $ \cite{ks}.
For larger $\xi_\alpha$,
instead, $T_\nu$ would be lower, since the neutrino decoupling 
temperature becomes larger than the muon mass \cite{ks}.
After neutrino decoupling, $L_\alpha$ and $T_{\nu}^3/s$ are constant,
and so are the $\xi_\alpha$.  

The energy density of stable relativistic 
neutrinos,
\begin{equation}
\rho = \sum_{\alpha = e,\mu,\tau} {\pi^2\over 15}  T_\nu^4
\left [ {7\over 8} + {15\over4\pi^2} \left( \xi_\alpha^2 + {\xi_\alpha^4\over
2\pi^2}\right )\right ]~,
\end{equation}
leads to an upper bound on the $\xi_\alpha$
 due to the limit on the present total
energy density of the Universe (in units of the critical density),  $\Omega_o$.
 The bound is
$\xi_e + \xi_\mu + \xi_\tau < 86$ for $ \Omega_o h^2 \lesssim
1/4$, as required if $\Omega_o \leq 1$ and $t_o \gsim 10^{10}$yr in a 
radiation dominated Universe \cite{ks}
 ($h$ is the Hubble constant in units of 100 km/sec Mpc, $h = 0.4
-1$ and $t_o$ is the present age of the universe). 
However, galaxy formation arguments provide a stronger upper limit.
Stable relativistic neutrinos in the large numbers considered here would 
maintain
the Universe radiation dominated (by the neutrinos) for much longer
(thus until lower temperatures)
than in the standard cosmology.  Requesting neutrinos to become subdominant
before the recombination epoch, Kang and Steigman \cite{ks} found, using 
nucleosynthesis bounds,
$-0.06 \lesssim \xi_{\nu_e} \lesssim 1.1$, $|\xi_{\nu_\mu,
 \nu_\tau}|  \lesssim 6.9$, $\eta = n_B/n_\gamma
  \lesssim 19 \times 10^{-10}$,
and, therefore, $\Omega_B h^2 =
\left (\eta /272.2 \times 10^{-10}\right ) \lesssim 0.069$. This
is an upper bound on $\Omega_B$ almost an order of magnitude larger than obtained 
in conventional nucleosynthesis. This would,
for example,  provide a solution
to what some call ``the x-ray cluster baryon crisis" \cite{white}
\cite{steigman}. In fact the conventional nucleosynthesis bound
on $\Omega_B$ combined with  the measurement of a ``large"
amount of gas in rich clusters of galaxies leads to an upper
bound on the amount of dark matter (DM) in the Universe, i.e.
$\Omega_{DM} \leq (0.2 -0.4) h^{-1/2}$ \cite{white}.
If $h > 0.16$, as all measurements confirm, this would mean
$\Omega_{DM} < 1$, namely that we either live in an open
Universe
or in a Universe with a cosmological constant. Since the upper
bound just mentioned on $\Omega_{DM}$ is proportional to
the nucleosynthesis upper bound on $\Omega_B$,
making the latter one order of magnitude larger would
eliminate this ``crisis". Notice that
 using Eq. (\ref{asymm}), the Kang and Steigman 
the bound $\xi_\alpha
\lesssim 6.9$ translates into $L_\alpha \lesssim 1.4$.

The production itself of a very large $L$ is problematic.
  In the standard out-of-equilibrium decay scenario for
the generation of the baryon asymmetry both $B$ and $L$ are
typically small.  The largest
asymmetries produced in this scenario are of order $L \simeq \epsilon (n_X
/g_{s*} n_\gamma)$, where $n_X$ is the number density of decaying
 particles and $\epsilon$ is the CP-violating
parameter that gives the lepton number generated per decay.  
Taking the generic value 
$g_{s*} \gsim 10^2$ at early times, we see that $\epsilon = 1$ and 
$n_x = n_\gamma$ are
necessary to get at most to $L \simeq 10^{-2}$.  Even if these values can be
arranged for, they are not easy to obtain in realistic models \cite{hk}. 
 Moreover, if the
baryon and lepton number violating (but $B-L$ conserving) reactions due to
sphalerons are in equilibrium after the production of an asymmetry $L$, they
will result in $B \simeq - L$, implying that the total $L$ must be as small as $B$.
Thus, if individual lepton numbers are large they should (almost) cancel each
other (with a fine tuning of eight orders of magnitude if $\xi \simeq O(1)$).

Here we present a viable model,  where naturally a large lepton number asymmetry can be
generated, $L\simeq (10^{-2} -1)$, corresponding to
$\xi_\nu \simeq O(1-10)$(see Eq. (\ref{asymm})), namely,
 much larger, by seven to ten orders of magnitude,
than the baryon asymmetry B.  
 We use the most efficient mechanism to produce a large fermion
number asymmetry, i.e. the decay of a scalar condensate carrying fermion number
in a supersymmetric model, as first proposed by 
Affleck and Dine in 1985 \cite{ad}.  Because
we want to generate only a large $L$ (not a large $B$ also),
 we consider models with a s-neutrino
condensate.

Most of the previous work dealing with the possibility of 
$L \gg B$
 either \cite{hk} \cite{lss} did not take into account the conversion of
  $L$ into $B$ due to
sphaleron mediated processes before the electroweak phase transition
 \cite{krs} (for temperatures $M_W/\alpha_w > T > M_W)$, or 
 \cite{ls} did not actually dealt with the
production of a large $L$. 
Only the following two potentially viable type of models have being so far proposed, to our knowledge, both very different from the
model we present here. One of them is due to
 Dolgov and Kirilova \cite{DK}. They pointed out that  a large $L$ and small $B$ could result within the Affleck and Dyne scenario,
 if very different rates of particle production (due to oscillations in 
field directions perpendicular to each flat direction) could be arranged for
the $L$ and $B$ carrying flat directions,  through a choice of coupling constants. Since particle production 
 depletes the  charge stored in the condensate, they suggested that
  the initially large
$B$ charge might be efficiently dumped by particle production and not the large 
$L$ charge. Secondly,
 models where a large $L$ asymmetry could be generated at low temperatures due to neutrino-antineutrino oscillations have been proposed \cite{FVS} recently. 

Our model is based on the Affleck and Dyne mechanism to generate a large $L$
and on the idea,
proposed by Linde twenty years ago \cite{l},
 that,  in the presence of a large enough lepton asymmetry
$L > L_C$, the electroweak symmetry is never restaured at
 large temperatures, and, therefore, 
 sphalerons are suppressed at all temperatures.  The necessary critical 
value $L_C$
depends on the standard Higgs field mass. For Higgs masses of 60 GeV to 1 TeV 
the critical values of $n_\nu/n_\gamma$ needed at $T > M_W$
range from  2.4 to 13.3 \cite{ls}. Using the relation 
$s = 1.80 g_{s*} n_\gamma$ and considering that a typical value of
$g_{s*}$ at $T > M_W$ is $\sim 100$, gives $L_C \simeq $(1.3 to
 7.4) $ 10^{-2} (100/ g_{s*} (T > M_W))$. 
Thus, if $L > L_C$ the weak gauge bosons are always massive and,
consequently, the rate of sphaleron reactions is always much smaller than the
rate of expansion of the Universe.  Therefore $B$ and $L$ would be preserved
 \cite{dmo}, or
at most a very small fraction of a large $L$ could be converted into $B$
by out of equilibrium sphaleron reactions,
giving origin to the small $B$ observed (if no larger $B$ was produced
earlier) \cite{ls}. Considering the bound from galaxy formation mentioned
above, we are interested in generating an $L$ in the range $L_C < L < 1.4$.

The particular particle model we study is the Minimal Supersymmetric Standard Model supplemented
with three electroweak singlet right handed neutrino superfields
$N_i$, $i = 1, 2, 3$ (whose scalar components will be denoted 
${\tilde N}_i$ in what follows) and supersymmetric masses
$M_i$.  We take the $N_i$ mass terms to be diagonal for simplicity.  Thus,
besides the terms of the MSSM, this model has in the superpotential
\begin{equation}
W = {1\over 2} M_i N_i^C N_i^C + h_{ij} N_i^c L_j H_u~.
\end{equation}
There is also a separate hidden inflationary sector,  not
leading to preheating \cite{kls}, that we do not need to
specify beyond giving the inflaton mass $m_\psi$ at the end of inflation and the inflaton decay rate $\Gamma_\psi$,
 for which we only use that the inflaton coupling
is gravitational (i.e. $g_\psi \simeq m_\psi/M_P$), 
thus $\Gamma_\psi \simeq m_\psi^3/M_P^2$ ($M_P$ is the
Planck mass, $1.2 \times 10^{19}$ GeV).  We assume the see-saw mechanism is
responsible
for the mass of the light (mostly left-handed) neutrinos, with hierarchical
right handed Majorana masses $M_1 < M_2 < M_3$ and Dirac neutrino masses
$m_{ij}^D \simeq h_{ij} v_u $ where $v_u = \langle H_u\rangle \simeq
10^2$ GeV (so that $M^D$ has eigenvalues $m_1^D$, $m_2^D$ and $m_3^D$ and,
if $\nu_1$ is the lightest and $\nu_3$ the heaviest neutrino, for example,
then $m_1^D < m_2^D < m_3^D $).

We assume that ${\tilde N}_1$ is a flat direction of the
potential during inflation. This implies that
$M_1$ is smaller than the expansion rate of the
Universe during inflation,  $M_1 < H_{inf}$.  In this case, 
through different possible mechanisms
that we will explore later, the field ${\tilde N}_1$ may find itself at the
end of inflation with a non-zero value, say ${\tilde N}_o$. 
Let us examine first how large
${\tilde N}_o$ needs to be, before studying the mechanisms that naturally can
lead to the range of values needed.  

The sequence of events we envision is the
following (see Fig.1).

\noindent $I$)-- The inflaton  $\psi$ 
starts oscillating about the true minimum
of its potential when the expansion rate of the Universe $H$ is 
$H_I \simeq m_\psi$.
This requirement comes from the solution of the evolution of a classical field
$\phi$ in an expanding Universe,
  $\ddot{\phi} + 3 H \dot{\phi} + \partial V/\partial\phi =
0$. Assuming $V(\phi) \simeq m^2 \phi^2$, the solution is oscillatory only for
$H < m$, while for $H > m$ the solution is overdamped and the field remains
stuck
at its initial value $\phi = \phi_o$.  The energy in the inflaton oscillations
redshifts like matter, so the universe becomes matter dominated (MD).

\noindent $II$)-- Through the same arguments the ${\tilde
N}_1$  sneutrino field oscillations start when $H= H_{II} \simeq M_1 < 
 H_I \simeq  m_\psi$. 
 We assume that at this point the
inflaton energy density still dominates, $\rho_\psi > \rho_{\tilde N}$. 
 
\noindent $III$)-- The $3^{rd}$ event of our sequence, the decay of the 
inflaton,  happens at a
later time $t \simeq \Gamma_\psi^{-1} \simeq M_P^2/m_\psi^3$. 
Using the MD relation
between H and t, we obtain that
 $H_{III} = (2/3) t^{-1} \simeq (2/3) \Gamma_\psi$. At this point the
Universe becomes radiation dominated (RD) by the decay products of the
inflaton. 

\noindent $IV$)-- The sneutrino ${\tilde N}_1$ decays later at 
$t \simeq \Gamma_{\tilde N}^{-1} \simeq
\left ( h^2 M_1/8\pi\right )^{-1}$ and using the RD relation between $H$ and
$t$, $H_{IV} = (2t)^{-1} \simeq h^2 M_1 /16\pi$. 
 Here $h$ is the largest $h_{1j}$ coupling.  This  $h$ 
is also the Yukawa coupling constant that dominates the Dirac
mass $m_1^D \simeq h v_u$.   Thus, using
$m_{\nu_1} \simeq h^2v_u^2/M_1$, we can eliminate the constant $h$ 
in favor of the lightest
neutrino mass in what follows.

\noindent $V$)--Thermalization happens when the rate of interaction of
particles in the Universe, $\Gamma_{int}$, becomes equal (and then larger) 
than the expansion rate of the Universe,  
i.e.  when $H = H_V = (2t_{th})^{-1} \simeq \Gamma_{int} 
\simeq n\sigma$, where $n$ is the number density of particles and
$\sigma$ their typical cross section.
The actual lepton asymmetry is either generated or released when the 
${\tilde N_1}$ sneutrino decays.
 We require to generate in total an asymmetry
$L > L_C\simeq 10^{-2}$.  It is important that the $\tilde N_1$ decay occurs
before thermalization is achieved.  If so,
thermalization happens in the presence of a large lepton asymmetry that breaks
the electroweak symmetry from the beginning of the thermal bath and,
consequently, the rate of sphalerons is always suppressed.

Let us examine the conditions for the previous sequence of events
to be consistent.
The measured anisotropy of the microwave background radiation requires
$H_{inf}$ to be  $10^{13}-10^{14}$ GeV in most models (even if it may be lower,
such as $10^{11}$ GeV, in some) \cite{HI}.  If the inflaton potential
during inflation is just quadratic (such as in chaotic inflation) then $m_\psi
\simeq {H_{\inf}/ 3}$ at the end of inflation (since $H_{\inf}^2 = (8\pi/3)
m^2_\psi \psi^2/M_P^2$ and generically  $\psi \simeq M_P)$.  However, with more
complicated potentials, $m_\psi$ can be smaller and still we can assume that
the potential at the end of inflation is well approximated by
 $ m_\psi^2 \psi^2$.  For
example if $V(\psi) = \lambda \psi^4 + m_\psi^2 \psi^2$, and the quartic term
dominates during inflation then $\lambda \simeq 10^{-13}$ \cite{HI}.
  Assuming that
$\psi\simeq M_P$ during inflation and
 $m_\psi\simeq (10^{-7}$ or $10^{-8})M_P$, the quadratic term
in the potential would dominate as soon as $\psi$ became smaller than $M_P$ 
(by a factor of 3 or 30), at the end of inflation. There are some models in 
which $m_\psi$ is even lower, $m_\psi= 10^{-9}M_P$ \cite{rs}. In what
 follows it will be important to have $m_\psi \lesssim 10^{12}$ GeV,
which is perfectly possible.

 Notice that $H_{III} \equiv H(t =
\Gamma_\psi^{-1}) \simeq 6$ MeV $(m_\psi/10^{12}$ GeV$)^3$ 
and we will be considering
values of $M_1$ larger than $H_{III}$, 
so that $\tilde N_1$ starts oscillating
before the inflaton decays $(H_{II} \simeq M_1 > H_{III}$). 
In our scenario $\tilde N_1$ decays after the inflaton decays 
(namely $H_{IV} < H_{III} )$ if $\Gamma_\psi/\Gamma_{\tilde N} >
1$. This  requires
\begin{equation}
M_1 < 1.6 \times 10^8 {\rm GeV} {(m_\psi/10^{12}{\rm GeV})^{3/2}\over
(m_{\nu_1}/10^{-4} {\rm eV})^{1/2}}~.
\end{equation}
The thermalization time and temperature,  $t_{th}$ and $T_{th}$,
 are obtained by estimating the number density of particles $n$
 and their cross section $\sigma$. We take $n \simeq P n_\psi$,
where $P$ is the
average number of daughter particles produced in a $\psi$-decay, and the
interaction cross section to be of electromagnetic order
$\sigma\simeq \alpha^2/E^2$. Here $E$ is the characteristic energy of the
particles, that redshifted from a value $E= m_\psi/P$ at 
$t = \Gamma_\psi^{-1}$,
thus $E = \simeq (m_\psi/P) [a(t = \Gamma_\psi^{-1}) /a(t_{th})]$.  Moreover,
$n_\psi$ redshifts from a value $\rho_\psi /m_\psi$ at 
$t = \Gamma_\psi^{-1}$, thus 
$n_\psi(t_{th})
= (\rho_\psi (t = \Gamma_\psi^{-1})/m_\psi) 
[a(t = \Gamma_\psi^{-1})/a(t_{th})]^3$
and $\rho_\psi(t = \Gamma_\psi^{-1})$ is the critical density at
$H = {2 \over 3} \Gamma_\psi$, i.e. $\rho_\psi (t = \Gamma_\psi^{-1}) \simeq
\Gamma_\psi^2 M_P^2/6\pi$.  We obtain $t_{th} = (36 \pi^2 M_P^2/\alpha^4 P^6
m_\psi^3)$. The essential condition $t_{th} > \Gamma_{\tilde N}^{-1}$ requires
(using $P=2)$
%
\begin{equation}
0.67 \times 10^4 {\rm GeV} {(m_\psi/10^{12}{\rm GeV})^{3/2} \over
(m_{\nu_1}/10^{-4} {\rm eV})^{1/2}} < M_1~.
\end{equation}
Notice that Eqs. (4) and (5) leave always an interval of four 
orders of magnitude for the allowed values of
$M_1$, independently of the values of $m_\psi$ and $m_{\nu_1}$.  Computing the
critical density $\rho_{th}$ at $t_{th}$, when $H = (2 t_{th})^{-1}$ and 
assuming that all that energy goes into radiation,
$\rho_{th} = (g_*/3) T_{th}^4$, we get the reheating temperature, that is
very low
%
\begin{equation}
T_{th} = {1\over 34}~~ {\alpha^2 P^3 m_\psi^{3/2}
\over g_*^{1/4} M_P^{1/2}}
 = 2.3\times 10^3 {\rm GeV} \left 
 ({\alpha\over 10^{-2} }\right )^2 \left ( {P\over
    2}\right )^3 {(m_\psi/10^{12}{\rm GeV})^{3/2} \over
(g_*/10^2)^{1/4} }~.
\end{equation}

We have not yet computed the asymmetry $L$.  As we will see the requirement of
a large enough 
$L$ sets a lower bound on $\tilde N_o$, the initial amplitude of the $\tilde
N_1$ oscillations.
We will initially consider the case when $\rho_\psi > \rho_{\tilde N}$
at $t= \Gamma_{\tilde N} ^{-1}$, which implies that
(the decay product of) the inflaton dominates the 
energy density at all previous times (see
Fig. 1). This sets an upper bound on $\tilde N_o$ and the viability of 
this variation of our model
depends on the existence on an allowed interval for $\tilde N_o$.

Defining $\epsilon$ as the net lepton number per $\tilde N_1$ particle
at decay, the total lepton  number density at $t=\Gamma_{\tilde
N_1}^{-1}$ is $n_L  \simeq \epsilon ~n_{\tilde N_1} (t =
\Gamma_{\tilde N_1}^{-1}) \simeq \epsilon~ \rho_{\tilde N} 
(t=\Gamma_{\tilde N}^{-1})/M_1$, where $\rho_{\tilde N}$
redshifted as matter from the moment ${\tilde N}_1$
oscillations started, i.e. $t=M_1^{-1}$, when $\rho_{\tilde N} = M_1^2 
{\tilde N}_o^2$ 
(recall that we assume the ${\tilde
N}_1$ potential is dominated by the quadratic term).  Thus,
%
\begin{equation}
n_L \simeq \epsilon~ M_1 \tilde N_o^2
\left [ {a(t = M_1^{-1})\over a (t = \Gamma^{-1}_{\tilde N} )}\right ]^3
= {\epsilon ~\tilde N_o^2 ~\Gamma_\psi^{1/2} ~\Gamma_{\tilde N}^{3/2} \over M_1}~.
\end{equation}
In obtaining Eq. (7) it is necessary to remember that the Universe goes from MD
to RD at $t = \Gamma_\psi^{-1}$, which occurs between $t = M_1^{-1}$ and $t =
\Gamma_{\tilde N}^{-1}$.  The entropy density at $t = \Gamma_{\tilde N}^{-1}$
is $s \simeq 4 \left [\rho_\psi (\Gamma_{\tilde N}^{-1})\right ]^{3/4}$
 and, since
$\rho_\psi$ dominates the energy density of the Universe, we can equal
$\rho_\psi$ to the critical density at the time, i.e.
 $\rho_\psi = (3/8 \pi) M_P^2 (\Gamma_{\tilde N}^2/4)$. We obtain
%
\begin{equation}
L = {n_L\over s} \simeq
{3.5 \epsilon \tilde N_o^2 \Gamma_\psi^{1/2}\over M_1~ M_P ^{3/2}}
\simeq {3.5 \epsilon {\tilde N}_o^2 m_\psi^{3/2} \over M_1~ M_P^{5/2} }~.
\end{equation}
The condition $L > L_C$ translates into
%
\begin{equation}
{\tilde N}_o > 0.54 ~10^{17}{\rm GeV}
\left ( {L_C\over 0.1~\epsilon}\right )^{1/2}
{(M_1/10^4 {\rm GeV})^{1/2}\over(m_\psi/10^{11} {\rm GeV})^{3/4} }~.
\end{equation}
On the other hand $\rho_\psi/\rho_{\tilde N} \simeq 0.03 (M_P/{\tilde N}_o)^2
(\Gamma_{\tilde N}/\Gamma_\psi)^{1/2}$ at $t = \Gamma_{\tilde N}^{-1}$, 
and requesting
$\rho_\psi/\rho_{\tilde N} > 1$ yields
%
\begin{equation}
{\tilde N}_o < 0.75 \times 10^{17} {\rm GeV}
 \left( m_{\nu_1}/ 10^{-4} {\rm eV}\right )^{1/4}
{(M_1/10^4 {\rm GeV})^{1/2}\over (m_\psi/10^{11} {\rm GeV})^{3/4}}~.
\end{equation}
This condition translates into an upper bound on $L$
\begin{equation}
L < 0.13 ~ \epsilon ~ (m_{\nu_1}/ 10^{-4} {\rm eV})^{1/2}  
\end{equation}
thus $L > L_C$ only if $\epsilon > 7 L_C (10^{-4}$ eV $/ m_{\nu_1} )^{1/2}$.
If  $m_{\nu_1} \simeq 10^{-4}$ eV, then $\epsilon$ needs to be near
maximum, $\epsilon \gsim 10^{-1}$. Let us see now that these
large values of epsilon are easily obtainable in our scenario.

There are two generic mechanisms to obtain a non-zero $\epsilon$,
i.e. a net lepton number per ${\tilde N}$ decay: one is to generate $\epsilon$
during the decay, due to CP and $L$ violation in ${\tilde N}$ decay modes,
starting from an $L = 0$, ${\tilde N}$ condensate, another is the generation of
an $L$ asymmetric condensate, due to CP and $L$ violating effective
operators.  In the MSSM model supplemented with three 
right-handed neutrinos ($N_i,\; i=1,2,3$)
the generation of $\epsilon$ in the ${\tilde N}$ decay has been studied
in the literature. If $M_3=M_1$, then
 $\epsilon \simeq ln(2) Im ( h_{33}^2 )/ 8 \pi$  \cite{msyy}
could be as large as $10^{-2}$, otherwise
there is a suppression factor  $\simeq M_1/M_3$ \cite{cdo}. 
Here $h_{33}$ is 
assumed to be the largest $h_{ij}$ coupling in Eq. (3), and 
$Im ( h_{33}^2 )$ could be as large as 1. However,
we prefer not to impose any lower bound on the
$\nu_3$ mass, thus leaving open the possibility of $M_3 >>
M_1$.  

This leaves us only the second method to account for the large
$\epsilon$ needed. In this case,
the decay of ${\tilde N_1}$, releases the
lepton-number asymmetry accumulated in the condensate 
due to the appearance of
effective lepton number violating soft supersymmetry breaking terms. These 
 operators  are due to the
usual mechanism of supersymmetry breaking 
in the MSSM, i.e. supergravity breaking
in a hidden sector.  The main term of this type in our model
 is $O \simeq m_{3/2} M_1 {\tilde N}_1
\tilde N_1$, which yields a lepton number per $\tilde N_1$ particle
$\epsilon = Im (<O>_o) ( M_1^2 {\tilde {N_o}}^2)^{-1}$ \cite{ad}, 
where $<O>_o  \simeq
m_{3/2} M_1 {\tilde {N_o}}^2$ is the initial value of the $L$-violating
operator, thus, assuming $Im (<O>_o) \simeq <O>_o$ we obtain
 $\epsilon \simeq m_{3/2}M_1^{-1}$.  
Replacing $m_{3/2} \simeq 1$ TeV, we see that $\epsilon \gsim 10^{-1}$
is possible if $M_1 < 10$ TeV, which
is allowed by Eq (5), even if $m_{\nu_1}\simeq 10^{-4}$ eV
(recall that $m_\psi$ can be  as low
as 10$^{11}$ GeV, a possibility in favor of which we argued above).
Actually, the value of $\epsilon$ obtained  is in general
higher than this estimate because at
the beginning of the oscillations the effective value of $m_{3/2}$
is $O(H)$, rather than $O(1$ TeV), due to the supersymmetry breaking
triggered by the non-zero vacuum energy (see below). Consequently,
$<O>_o$ is much larger, an thus the $L$ generated. Actually, a value
$\epsilon\simeq O(1)$ may be available, even for $M_1 < 10$ TeV. So,
the previous estimate  $\epsilon \simeq O(1$ TeV$)M_1^{-1}$, represents
a lower bound on $\epsilon$ rather than a precise value.

We see in Eq. (11) that even with $\epsilon \simeq 1$, $L$ can be 
up to $O(10)~L_C$
but not much larger, unless $m_{\nu_1} > 10^{-4}$ eV. We have so far kept
$m_{\nu_1} \simeq 10^{-4}$ eV because this value
can easily accommodate the MSW solution to the solar neutrino problem. 
Since the mass square
difference between $\nu_1$ and $\nu_2$ needs to be 
$\Delta m^2 \simeq 10^{-6}$ eV$^2$
it is enough to choose $m_{\nu_1} < m_{\nu_2} \simeq 10^{-3}$ eV.  However
many models including this solution, mainly those trying
to accommodate simultaneously various of the hints  for non-zero neutrino 
masses \cite{neut}, propose much larger $m_\nu$ values, such as
$m_{\nu_1} \simeq O($eV). In this case  the two oscillating
neutrinos are almost degenerate, $m_{\nu_1} \simeq m_{\nu_2}$. 
 This is the scenario we would need to accept in order to
relax numerically the bound from Eq. (10) by one order of magnitude,
i.e. for $m_{\nu_1} = 1$ eV, keeping $M_1$ and $m_\psi$ the same,
we get from Eq.(10)
%
\begin{equation}
{\tilde N}_o < 0.75 \times 10^{18} {\rm GeV}
 \left( m_{\nu_1}/ 1 {\rm  eV}\right )^{1/4}
{(M_1/10^4 {\rm GeV})^{1/2}\over (m_\psi/10^{11} {\rm GeV})^{3/4}}~.
\end{equation}
In this case, from Eq (11) we see that $L<13 \epsilon$
with $m_{\nu_1} \simeq 1$ eV, so $L$ could easily be up to
$10^2 L_C$. 

We see that the best solutions, those
with larger $L$, tend to saturate the upper 
bounds in Eqs.(10) and (11) so that at 
$t =\Gamma_{\tilde N}^{-1}$ we have $\rho_{\tilde N} \simeq \rho_\psi$.
This leads us to the second variation of our model, namely to relax the
condition of $\psi$-dominance, that yield Eqs. (10) or (12),
and allow $\rho_{\tilde N}$ to become larger than
$\rho_\psi$  in the period, after $t = M_1^{-1}$, 
in which $\rho_{\tilde N}$ behaves like
matter ($\tilde N_1$ is oscillating) and $\rho_\psi$ (actually the 
density of their decay products) behaves like radiation.

Requesting the moment of equality, $t_{eq}$,
namely the moment  at which $\rho_{\tilde N} (t_{eq})$
 = $\rho_{critical}$, to happen before the ${\tilde N}$ decay,
 that is $t_{eq} \simeq 0.5 \left (M_P/{\tilde N}_o\right)^4
\Gamma_\psi^{-1}  < \Gamma_{\tilde
N}^{-1}$, reverses the inequality in Eq. (10) (and Eq. (12)), 
into an upper bound on ${\tilde N}_o$
%
\begin{equation}
{\tilde N}_o > 0.75 \times 10^{17} {\rm GeV} \left ({m_{\nu_1}\over
10^{-4}{\rm eV}}\right)^{1/4}
{(M/10^4 {\rm GeV})\over (m_\psi/10^{11} {\rm GeV})}^{3/4}~.
\end{equation}

In this case the s-neutrino decay is responsible for both, the production of
the lepton asymmetry in fermions and the reheating of the Universe.
Since $n_L = \epsilon \rho_{\tilde N} (t=\Gamma_{\tilde N}^{-1}) /M_1$ and 
 $s=4  \left [
\rho_{\tilde N} (t=\Gamma_{\tilde N}^{-1})\right ]^{3/4}$, we obtain
%
\begin{equation}
L = {n_L\over s} \simeq {\epsilon\over 4 M_1} \left ( \rho_{\tilde N}
(\Gamma_N^{-1})\right )^{1/4} \simeq 0.1 \epsilon {\sqrt{M_P\Gamma_{\tilde N}}
\over M_1}~.
\end{equation}
In obtaining this equation we need to notice that, while
 at the beginning of the ${\tilde N_1}$ oscillations we still have
$\rho_{\tilde N} = M^2{\tilde N}_o^2$, as before (see Eq. (7)), 
now in the redshift factor
$\left [a(t = M^{-1}) / a(t = \Gamma_{\tilde N}^{-1} )\right ]^3$
we have to take into account that the Universe is radiation dominated between
$t \simeq \Gamma_\psi^{-1}$ and $t_{eq}$, 
and matter dominated before and after (until
${\tilde N_1}$ decay), see Fig. 1.  Substituting,
 $\Gamma_{\tilde N} \simeq  m_{\nu_1}
M_1^2/8\pi v_u^2$ we get
%
\begin{equation}
L = 0.2 ~\epsilon \left ( {m_{\nu_1}\over 10^{-4}{\rm eV}}\right )^{1/2}.
\end{equation}
This $L$ is about the maximum allowed in the
previous scenario, Eq (11).
Eq. (15) shows that $L$ can be larger for larger values of $m_{\nu_1}$.
On the other hand, too large values of $m_{\nu_1}$ may lead to erasure of $L$. 
In order to see this point,
let us consider the thermalization in this scenario. Here, 
the thermalization occurs instantaneously after the ${\tilde N}$ decay,
 because the rate of  interaction of
the decay products $\Gamma_{int} = P n_{\tilde N} \sigma = P^3 n_{\tilde N}
\alpha^2 M_1^{-2}$ at the moment of decay, is larger than the expansion rate 
of the Universe at that time, $H\simeq \Gamma_{\tilde N}$,
%
\begin{equation}
{\Gamma_{int} \over \Gamma_{\tilde N}} \simeq {\alpha^2\over 48\pi^2}~
 P^3 {M_P^2\over v_u^2} {m_{\nu_1}\over M_1} \simeq 2 \times 10^{11} \left
({\alpha\over 100} \right )^2 \left ({P\over 2} \right )^3 \left ( {m_{\nu_1}
\over 10^{-4} {\rm eV} } \right )
\left ( {10^4 {\rm GeV}\over M_1} \right )~.
\end{equation}

Thus, the ${\tilde N}$ energy density is immediately thermalized after the
decay,
$\rho_{\tilde N} (\Gamma_{\tilde N}^{-1}) = (3/8\pi) M_P^2 \Gamma_{\tilde N} ^2
= (g_*/3) T_{th} ^4$, and
%
\begin{equation}
T_{th} \simeq {{M_1\sqrt {M_P~m_{\nu_1}}\over(30 \pi^3 g_*)^{1/4} v_u}}\simeq
{5\times 10^3 {\rm GeV}\over (g_*/100)^{1/4} }
\left ({M_1\over 10^4 {\rm GeV} }\right )
 \left ({m_{\nu_1}\over 10^{-4} {\rm  eV}} \right )^{1/2}~.
\end{equation}
Notice that $T_{th} < M_1$ if
$m_{\nu_1} < 4 \times 10^{-4}$ eV, 
thus the sneutrinos ${\tilde N_1}$ are not present in the thermal bath. 
For larger values of $m_{\nu_1}$,  $T_{th} > M_1$ and  one should
worry about the erasure of $L$ caused by the decay of the ${\tilde N_1}$
 present in the
thermal bath. Even if the $L$ initially generated could be as large as
 $10^2 L_C$, with $m_{\nu_1} \simeq 1$ eV this erasure could lower $L$
 considerably.
  If we do not want to worry about this source of $L$-erasure, the value
of $m_{\nu_1}$ is fixed at $m_{\nu_1} \simeq 10^{-4}$ eV, since it cannot be
smaller for $L \gsim L_C$, Eq. (15).

Notice that in the first variation of our 
scenario the preferred neutrino masses
are of order eV or larger. Without
any major modification of our model we
could choose $\nu_1$ to be heaviest of the
three light neutrinos. In this case
the lightest right handed neutrino would
mix in the mass matrix preferably with
the heaviest light neutrino and not the
lightest, in a see-saw scenario with
$m_{\nu_1}^D \simeq m_{\nu_2}^D \simeq m_{\nu_3}^D$.
Doing so the two lightest neutrino masses
are unconstrained in our model.
We could do the same thing in the second
variation of our scenario, but in this
case the preferred value of the heaviest neutrino
would be $10^{-4}$ eV. This would not be
compatible with the heaviest neutrino accounting
for part of the dark matter, for example
something totally possible in the first variation.

Let us address the issue of how the necessary $\tilde {N_0}$ could be generated.
We have seen that in the first scenario we considered that $\tilde {N_0}$
has to be between $10^{17}$ and $10^{18}$ GeV, 
while in the second scenario  we need $\tilde {N_0} > 10^{17}$ GeV.
There are several mechanisms that can produce a value of ${\tilde N}_o \simeq
10^{-2} M_P$ or larger. 
 Let us recall that during inflation supersymmetry 
  is necessarily broken
since the energy density $V_o$ is non zero.
This induces  a mass 
proportional to the Hubble parameter $H$ for all flat
directions, thus $m_{\tilde N_1}^2
= c~ H^2 + M^2$. 
Dine et al. \cite{drt} 
assumed that $c$ is negative and of order one, and that
non-renormalizable terms $\delta W \simeq \lambda
\tilde N_1^n / ( n (\beta M_P)^{n-3})$ 
in the superpotential stabilize the minimum
to the value
%
\begin{equation}
\tilde N_o = 
\left ( {\sqrt{-c}} ~ H (\beta M_P) ^{n-3}\over (n-1) ~ \lambda
\right ) ^{1/ (n-2)}~.
\end{equation}
It is obvious that $H < \tilde N_o < M_P$ (with the constants $\beta$ and $c$ of
order one) and $\tilde N_o$ becomes arbitrarily closer to $\beta M_P$ for large
$n$. The conditions to obtain the necessary negative effective masses and
non-renormalizable terms to lift the ``flat" direction of the potential
in supergravity scenarios have in general been examined in 
Ref. \cite{cg}, both in D- and F- inflationary models. Large negative
masses square favour F-inflationary scenarios in which the K{\"a}hler
potential contains a large mixing of the sneutrino and the inflaton in
 the term quadratic in $N_1$. This is a strong constraint on supergravity
 scenarios, which, for example, excludes minimal supergravity and in 
 string-based models requires that the inflaton be a $T$ modulus field
 (and not the dilaton $S$). In particular, in orbifold constructions,
 $m_{\tilde N_1}^2 \leq 0$ at the origin, is obtained  if the modular 
 weight of the $N_1$ field is  $n_{\tilde N_1} \leq -3$.

There is another possibility we think cannot be neglected, namely that $c$ is
 very small. The value $m_{\tilde N_1}^2 \simeq 0$ at the origin 
 can also appear in a natural
way, with no fine-tuning at all, both in $F$-inflation and in $D$-inflationary
scenarios \cite{cg}.
  In this case, $\tilde N_o$ departs from zero due to
quantum fluctuations during the Sitter phase.  The coherence length of these
fluctuations $\ell_{coh} \simeq H_{inf}^{-1} \exp (3 H_{inf}^2/2m^2) \simeq
H_{inf} ^{-1} \exp (3/2c)$ would be large enough to encompass our present
visible Universe if $c < 2 \times 10^{-2}$. When the coherence
length is much larger than the horizon, $\ell_{coh} \gg~ H^{-1}_{inf}$, 
as is the case here, the field
$\tilde N_1$ can be treated as classical and
%
\begin{equation}
\tilde N_o = {\sqrt{\langle \tilde N_1^2\rangle}} \simeq {\sqrt{3\over 8\pi^2}
\quad {H_{inf}^2 \over M}} = \sqrt {3\over 8\pi^2}\quad {H_{inf} 
\over{\sqrt c}}~.
\end{equation}
For $\tilde N_o \simeq 10^{17}$ GeV and $H_{\inf}~ \simeq ~10^{13}$ GeV we need
$c\simeq 4 \times 10^{-10}$.
This small value of $c$ can be naturally achieved in string-based
$F$-inflationary  models.
 In $T$-driven  models with orbifold 
compactifications this possibility is insured by the discrete character
of the modular weights, since it is enough that  $n_{\tilde N_1}=-3$. 
In $D$-inflationary models,
usually the $D$-term is associated to an anomalous U(1), and 
$m_{\tilde N_1}^2 \simeq 0$ results if the $N_1$ field is a singlet under
that group \cite{cg}.

In conclusion, we presented here a model to produce a very large lepton
asymmetry $L\simeq 10^{-2} - 1$ without producing a large baryon asymmetry.
The model is based on the Affleck and Dyne scenario, in which the field
acquiring a large vacuum expectation value during inflation is an sneutrino
right. We take as a specific model the MSSM supplemented with three right
handed neutrino singlet superfields, besides an inflationary sector.
 We considered two variations of one scenario, one in which the inflaton
energy dominates at reheating and another in which the sneutrino energy 
dominates then. In both cases we required $L$ to be large enough for the
 electroweak symmetry  to be spontaneously broken at all temperatures after 
 inflation, with the consequent suppression of sphaleron transitions. 
 In order to obtain this, in our scenario the large $L$  
 is generated before thermalization. The models considered are 
realistic. The first variation works better if the 
lightest neutrino mass is of $O(1$ eV). 
 In this case the MSW solution 
to the solar neutrino problem would require almost degenerate neutrinos,
as proposed in models trying to account simultaneously for several of the 
present hints for non-zero neutrinos masses.  Alternatively, if the
bounds apply to the heaviest neutrino instead, the lighter neutrino
masses are unconstrained.
The second variation works better when the
lightest neutrino mass is of $O( 10^{-4})$ eV, what can easily accommodate the 
masses needed for the MSW mechanism. In both cases the preferred value
of the lightest right handed neutrino is of  $O($TeV) and the vacuum
expectation value of the s-neutrino field during inflation must be larger
than $10^{17}$ GeV. We also commented on ways to obtain naturally these large
values for the sneutrino condensate. 
We have not addressed here how the baryon asymmetry could be generated. 
 A very interesting possibility that deserved further study is that the 
 out of equilibrium sphaleron transitions may translate a minor part of the 
 $L$ asymmetry into $B$ \cite{ls}.

\section*{Acknowledgements}
We thank the Aspen Center for Physics, where this paper was 
initiated, for its hospitality. 




\pagebreak
\begin{figure}[p]
\centerline{\epsffile{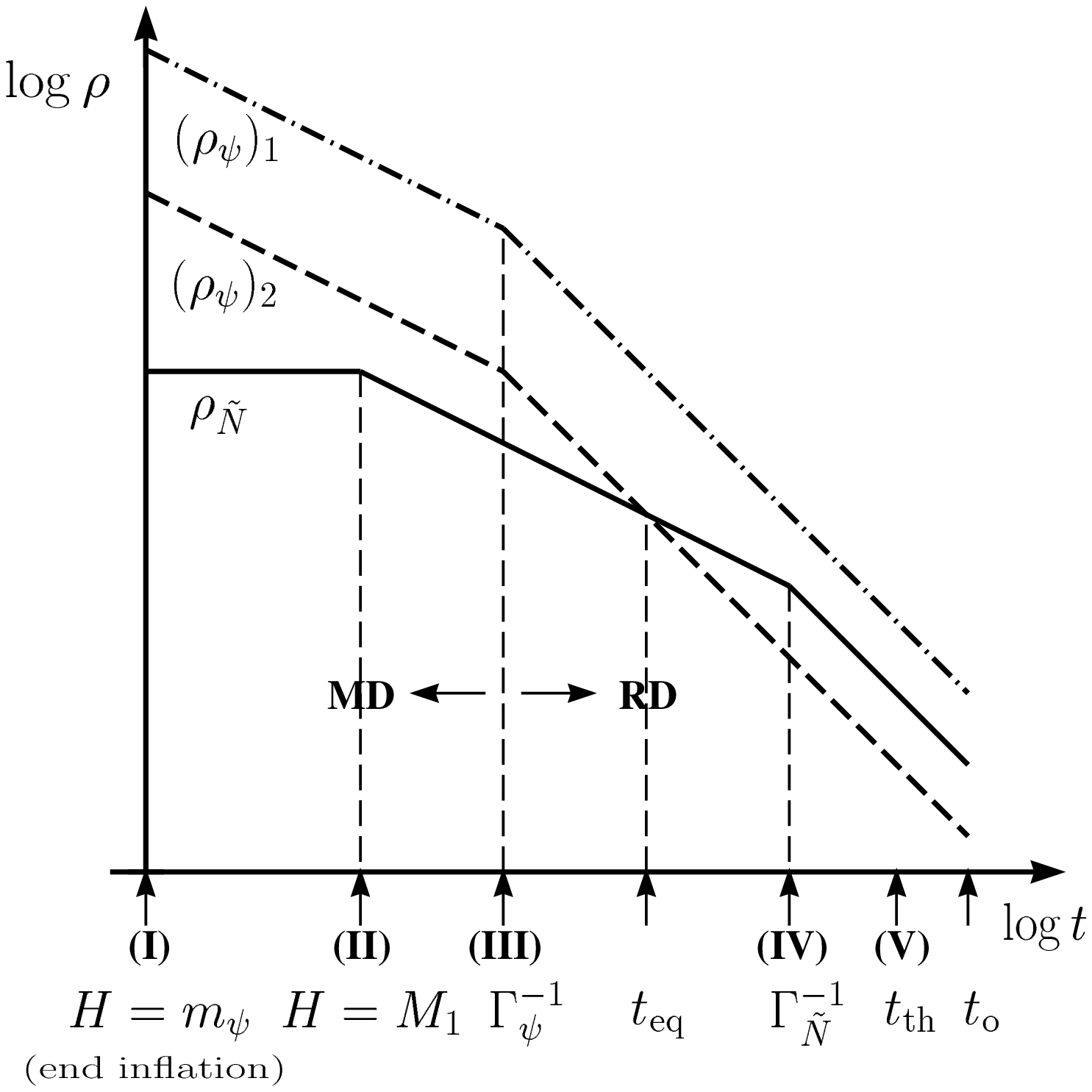}}
\caption{Schematic view of the sequence of events we envision after the end of 
inflation. See the text for details.
 The dashed-dotted  and dashed lines show the energy density of the 
inflaton (before III) and its decay products (after III)
 in the first and second
variations of our scenario, respectively.
The solid line shows the energy density of the relevant
sneutrino and its decay product. The Universe is matter dominated
(MD) before the inflaton decays (III) and in the first variation it is
radiation dominated (RD) afterwards. In the second variation de 
Universe becomes again MD for a period before the sneutrino decays.}
\end{figure}

\end{document}